\documentclass{eas}
\usepackage{graphicx}
\newcommand{\apj}{\mbox{\it Astrophys. J.}}
\newcommand{\apjl}{\mbox{\it Astrophys. J.}}
\newcommand{\apjs}{\mbox{\it Astrophys. J.}}
\newcommand{\aap}{\mbox{\it Astron. Astrophys.}}

\newcommand{\mnras}{\mbox{\it Mon. Not. R. Astron. Soc.}}
\newcommand{\nat}{\mbox{\it Nature}}

\newcommand{\physrep}{\mbox{\it Phys. Rep.}}

\newcommand{\beq}{\begin{equation}}
\newcommand{\eeq}{\end{equation}}
\newcommand{\ba}{\begin{array}}
\newcommand{\ea}{\end{array}}

\def \etal{{\it et al.~}}

\def\gtsima{$\; \buildrel > \over \sim \;$}
\def\ltsima{$\; \buildrel < \over \sim \;$}

\def\gsim{\lower.5ex\hbox{\gtsima}}
\def\lsim{\lower.5ex\hbox{\ltsima}}

\begin{document}

%%-----------------------------
%%      the top matter
%%-----------------------------
\title{Radiative Mechanisms in GRB prompt emission} 
\author{Asaf Pe'er}\address{Physics Department, University College Cork, Cork, Ireland}

\begin{abstract}

Motivated by {\it the Fermi gamma-ray space telescope} results, in
recent years immense efforts were given to understanding the mechanism
that leads to the prompt emission observed. The failure of the
optically thin emission models (synchrotron and synchrotron self
Compton) increased interest in alternative models. Optically thick
models, while having several advantages, also face difficulty in
capturing several key observables. Theoretical efforts are focused in
two main directions: (1) mechanisms that act to broaden the Planck
spectrum; and (2) combining the optically thin and optically thick
models to a hybrid model that could explain the key observables.

\end{abstract}
\maketitle
%%-----------------------------
%%      your text
%%-----------------------------

\section{Setting the stage: understanding what we see}
\label{sec:intro}

In the commonly accepted gamma-ray bursts (GRB) ``fireball'' model 
[ \cite{Pac86, Goodman86, SP90, RM92, RM94}], the prompt emission is
believed to arise from a prompt dissipation of a substantial fraction
of the bulk kinetic energy of a relativistic outflow, originating from
a central compact object. This model is found to be in good
qualitative agreement with all observations to date; moreover, a great
success of this model is the prediction of the afterglow emission,
resulting from interaction of the propagating relativistic blast wave
with the ambient interstellar matter (ISM) [\cite{MR97, SPN98}].

In spite of these successes, this model is far from being
complete. Many necessary details are missing: for example, the
mechanism responsible for particle acceleration to high energies,
required to explain the observed high-energy non-thermal emission is
not explained. Similarly, the nature of the radiative processes that
produce the observed signal are not specified. In addition, the
dynamical part is not fully understood. While it was long thought that
the conversion of explosion (gravitational) energy to kinetic energy
(namely, acceleration to relativistic velocities) is mediated by
photons [\cite{Pac86, Pac90, RM92}], in recent years there are
accumulating evidence that magnetic field may play an important role
in this process [\cite{ZP09}], resulting in a modified dynamics
[\cite{Drenkhahn02, DS02}]. Moreover, nothing in the model predicts
the radii in which energy is dissipated and radiation is produced.

The prompt GRB spectra is well modeled by a smoothly broken power law,
known as the ``Band'' function [\cite{Band+93, Preece+98a, Preece+00,
  Kaneko+06, Nava+11a, Goldstein+12}]. In spite of its great success in
providing good fits to the observed data, this model has a crucial
drawback: being mathematical in nature, by itself it does not provide
any clue about the origin of the observed emission.

It was long thought that the observed radiation originates from
synchrotron emission in the optically thin regime [\cite{MLR93, MRP94,
  Tavani96, Cohen+97}]. This idea was motivated by the fact that the
observed radiation is non-thermal. Shock waves which are
believed to exist in the plasma can accelerate particles to high
energies via Fermi mechanism as well as generate strong magnetic
fields, thereby providing the necessary ingredients for synchrotron
emission [\cite{BE87}]. These processes were recently realized in
particle-in-cell (PIC) simulations [\cite{Spit08}, \cite{SS09, SS11, Haug11}].

Although GRB spectra significantly vary from burst to burst and frequently
within a single burst, there are several key observations which appear
general. The synchrotron theory can therefore be confronted with these
key results. These include: 
\begin{enumerate}
\item Observed peak energy $E_{\rm peak}^{ob} \sim 300$~keV. While the
  synchrotron theory does not naturally provide this value, it is
  achievable under the assumption that both the electrons and magnetic
  field energies are close to equipartition with the post-shock
  thermal energy. For example, if the magnetic field is $B \approx
  10^5$~G, the characteristic electron's Lorentz factor is
  $\gamma_{el} \sim 200 $ and bulk Lorentz factor $\Gamma \sim
  10^{2.5}$, similar values are obtained.
\item Narrow distribution of the peak energy: although the observed
  luminosity varies by several orders of magnitude, in most GRBs the
  observed peak energy is between $0.1 - 1$~MeV. In the context of the
  synchrotron model, the observed peak energy is a function of $B$,
  $\gamma_{el}$ and $\Gamma$.  There is no natural reason to assume that
  the values of these free model parameters coincide in such a way as
  to produce the narrow clustering of $E_{\rm peak}^{ob}$ observed.
\item The correlation seen between the peak energy and total energy
  ($E_{\rm peak} - E_{\rm iso}$ relation) [\cite{Golenetskii+83,
    Amati+02, Ghirlanda+04, Yonetoku+04}]: in the framework of the
  synchrotron model, it is possible to obtain the observed correlation
  only if additional assumptions are made, e.g., about the dissipation
  radius.
\item A 'universal' low energy spectral slope, $\alpha \approx -1$
  [\cite{Kaneko+06, Nava+11a, Goldstein+12}]: in the ``Band'' model
  fits, a narrow clustering of the low energy spectral slope ($dN/dE
  \propto E^{\alpha}$) around $\alpha \approx -1$ is observed.  The observed
  low energy hard spectral slope is in contradiction to the prediction
  of the synchrotron model theory.  This is known as 'synchrotron
  (model) line of death' [\cite{Preece+98, Preece+02, GCG03}].
\end{enumerate}

The failure of the synchrotron model has motivated the study of
alternatives. A notable alternative is emission from the {\it
  optically thick} regions. While many of the details of the
``fireball'' model are uncertain, the existence of an optically thick
region in the inner parts of the outflow is a robust prediction. Thus,
photospheric emission is a natural outcome of the model, and, indeed
was considered from the very early days [\cite{Goodman86,
  Pac86}]. However, as the observed spectrum {\it does not} resemble a
Planck spectrum, this idea was abandoned for a long time.

\section{Broadening mechanisms of Planck spectrum: sub photospheric energy dissipation}
\label{sec:broadening}

The observed low energy spectrum is steeper than synchrotron model
predictions, but is not as steep as to resemble a ``Planck''
spectrum. However, while there is no physical mechanism that can
steepen the synchrotron spectra, one can think of several mechanisms
that can broaden the Planck spectrum to produce the observed spectral
slope.

Broadly speaking, there can be three ways in which the observed
spectra can be achieved. First, the spectrum may contain two separate
components: a ``Planck'' and optically thin synchrotron observed
simultaneously. The observed spectrum is a combination of these two
components. Following early analysis by \cite{Ryde04, Ryde05} and
\cite{RP09}, recently, with improved {\it Fermi} capabilities that
enable {\it time-resolved} analysis, these components are ubiquitously
observed [\cite{Ryde+10, Guiriec+11, Larsson+11, Zhang+11,
    axelsson+12, Starling+12, Guiriec+12}]. The separation enables the
study of the physical properties of both components [\cite{Peer+07,
    ZLB11, Peer+12}], and provides a natural explanation to the delay
of the high energy emission seen [\cite{Abdo+09a, Ackermann+10}].

Second, sub-photospheric energy dissipation naturally leads to
modification of the Planck spectrum [\cite{PMR05, PMR06, Giannios06,
    Ioka+07, Giannios08, Lazzati+09, Beloborodov10, LB10, Vurm+11,
    Giannios12}]. The basic idea is that kinetic energy dissipation,
whether originating from internal shocks, magnetic reconnection or any
other process, takes place at radii not much below the photospheric
radius. By definition of the photospheric radius $r_{ph}$, the optical
depth for scattering of a photon from $r_{ph}$ to the observer
(located at infinity) is equal to unity. The plasma contains many more
photons than electrons: this can be seen by the fact that the average
energy per photon (in the comoving frame) is much smaller than $m_e
c^2$. Thus, while at $r_{ph}$ the optical depth for {\it photon} scattering
is unity, the optical depth for {\it electron} scattering is much
larger than unity. As a result, at $r_{ph}$, every electron undergoes many
inverse Compton (IC) scatterings with the lower energy photons before
decoupling. Each electron therefore loses its energy rapidly, on a
time scale much shorter than the dynamical (expansion) time scale (see
\cite{PMR05} for details).

Assuming that the heating mechanism (of an unspecified nature) is
continuously heating the electrons, or alternatively accelerating new
electrons to high energies, the result is that the electron's
distribution is in a quasi steady state, with temperature determined
by balance between the external heating and the rapid IC cooling. This
temperature is inevitably higher than the photon temperature, hence
the plasma is characterized by {\bf two temperatures}: $T_{el} >
T_{ph}$.

If the dissipation, hence the electron heating occurs below, or even
slightly above the photosphere, then the thermal photons IC scatter
with the hotter electrons, producing a non-thermal spectrum.  The
emerging spectrum above the original thermal peak depends mainly on
two free model parameters: (1) the optical depth $\tau$ in which the
dissipation takes place: this determines the number of scattering for
a single photon. On the one extreme, $\tau \rightarrow \infty$ (or
$r_{dis} \ll r_{ph}$), the plasma have enough time to thermalize, and
the energy given to the electrons is evenly distributed, resulting in
a Planck spectrum. On the other extreme, $\tau \ll 1$, only very few
photons are being up scattered, producing a high energy tail. In the
intermediate regime, $\tau \approx$~few - few tens, the spectrum
significantly deviates from Planck. (2) The second free parameter is
the ratio of the energy density in the electron and thermal photon
components. If the dissipation considerably heats the electrons,
deviation from a Planck spectrum is more pronounced.

\begin{figure}
\includegraphics[width=10 cm]{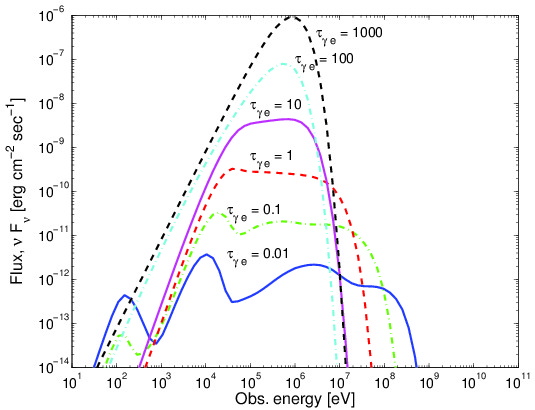}
\caption{Examples of time averaged spectra obtained for different
  values of the optical depth for photon scattering ($\tau =
  \tau_{\gamma e}$) at the dissipation radius, under the assumption
  that thermal component exists (from \cite{PMR06}). While for very
  high optical depth a ``Planck'' spectrum is obtained, for intermediate
  optical depth, multiple IC scattering results in nearly flat spectra
  above the thermal peak, while synchrotron emission modifies the
  spectrum at lower energies. }
\label{fig:1}
\end{figure}

Multiple IC scattering thus modifies the spectrum above the thermal
peak. At lower frequencies, the spectrum is dominated by synchrotron
emission from the energetic electrons. As these electrons are in a
quasi steady state, the emerging spectrum does not expect to have a
power law shape, as the electrons distribution cannot be described by
a power law. Thus, overall, the expected spectra is expected to
significantly deviate from the original Planck spectra, with
significant synchrotron contribution at low energies, and high energy
spectra dominated by multiple IC scatterings. Example of possible
spectra under different conditions appear in Figure \ref{fig:1}, taken
from \cite{PMR06}. Recently, evidence for sub-photospheric energy
dissipation was observed in analyzing the data of GRB090902B
[\cite{Ryde+11}].

\section{Theory of photospheric emission from collimated outflow}
\label{sec:geometry}

Even in the absence of sub-photospheric energy dissipation, the
expected spectrum originating from the photosphere deviates from a
pure ``Planck'' spectrum.  This is due to the non-trivial shape of the
photosphere. Consider first a spherical explosion: the mean free path of
photons emitted from high angle to the line of sight, $\theta > 0$ and
propagate towards the observer is larger than the mean free path of
photons propagating at $\theta = 0$. This results in a strong angular
dependence of the photospheric radius, $r_{ph} \propto \Gamma^{-2} +
\theta^2/3$ [\cite{Peer08}], where $\Gamma$ is the bulk Lorentz factor.

Moreover, by definition, the photospheric radius is the radius in
which the optical depth for scattering $\tau = 1$. However, the last
scattering process is not limited to this surface: photons have a
finite probability of being scattered at any position in space in
which scatterers (electrons) exist. An observer therefore sees
simultaneously photons who's last scattering location took place at a
range of radii and angles to the line of sight; this leads to the
concept of a ``fuzzy photosphere''. As photons adiabatically cool
below the photosphere, each of the observed photons has its own
(comoving) energy, and has a unique Doppler boost. Thus, the observed
spectrum differs than Planck spectrum, and is observed as a ``gray
body'' spectrum [\cite{Peer08, Beloborodov11}]. If one considers a
$\delta$-function of emission at $t=0$ (alternatively, if the inner
engine is abruptly stopped), then at late times emission is dominated
by photons emitted at high angles (off-axis). In this scenario, a very
flat spectrum is obtained at late times, significantly different than
a ``Planck'' [\cite{PR11}].

While the original theory was developed for spherical outflows, in any
realistic scenario the explosion is collimated. In the collapsar
model, for example, as the jet drills its way through the collapsing
stellar envelope it pushes material towards the side, forming a hot
cocoon. This material collimates the jet [\cite{ZWM03, Morsony+07,
    Mizuta+11}]. Thus, when calculating emission from the photosphere
one needs to consider the jet velocity profile, $\Gamma =
\Gamma(\theta)$. Such a model therefore has 4 free parameters (as
opposed to a single parameter, $\Gamma$ in the spherical case): the
maximum bulk Lorentz factor $\Gamma_0$ at the jet axis, the
characteristic jet opening angle $\theta_j$, viewing angle $\theta_v$
and a parameter $p$ which determines the shape of the velocity profile
decay ($\Gamma(\theta) \propto \theta^{-p}$).

\begin{figure}
\includegraphics[width=\linewidth]{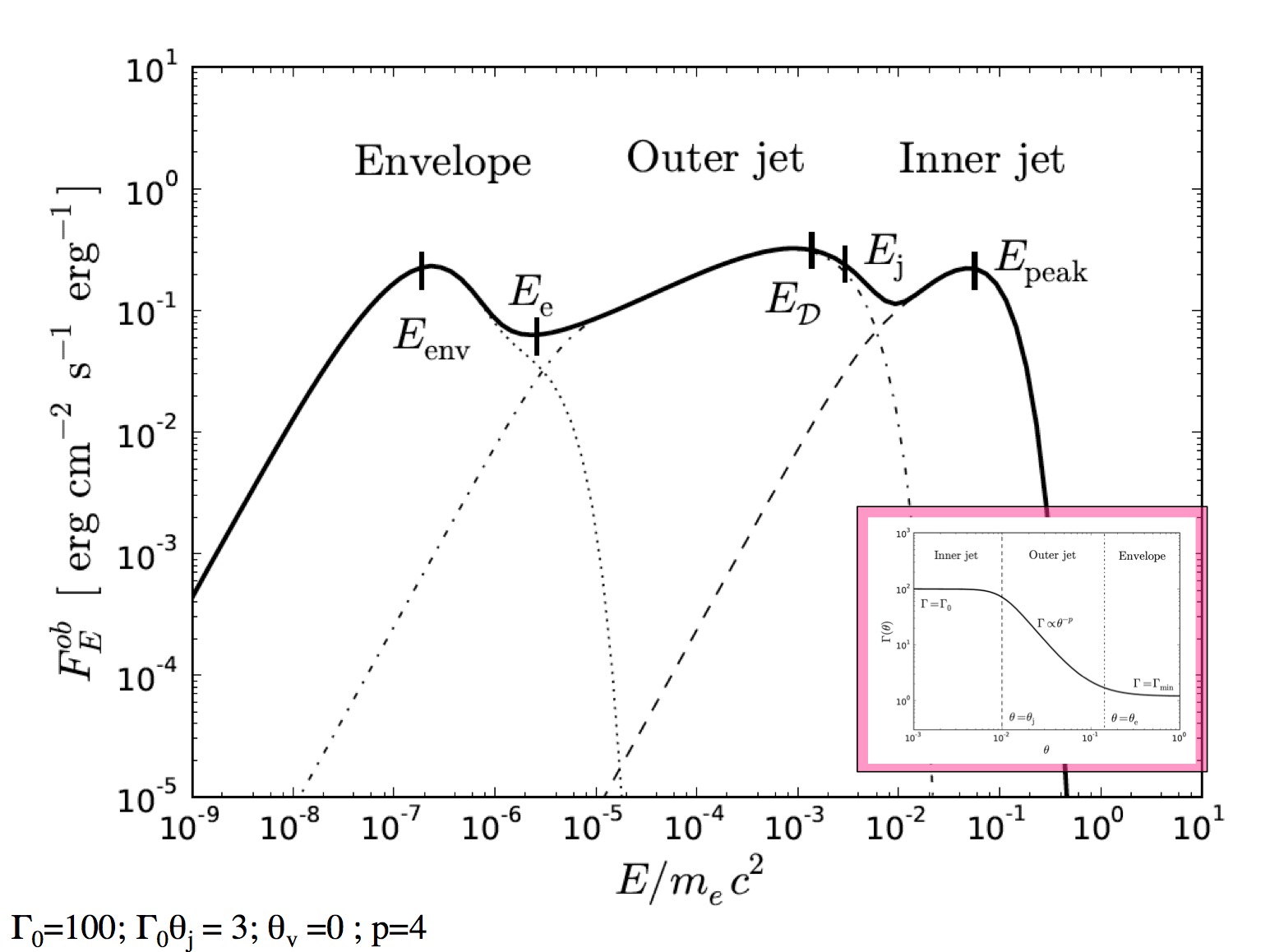}
\caption{Example of observed spectrum from a relativistic, optically
  thick outflow (taken from \cite{LPR12}). A jet profile
  $\Gamma(\theta) \propto \Gamma_0 /(\theta/\theta_j)^{2p}+1)^{1/2}$
  was considered (inner onset). Separate contributions from the inner
  jet ($\Gamma \approx const$), outer jet ($\Gamma \propto
  \theta^{-p}$) and envelope ($\Gamma \gsim 1$) are marked. The
  combined effect is a very flat spectra, extended over many orders of
  magnitude. This result is found to be robust, very weakly dependent
  on the values of the free model parameters }
\label{fig:2}
\end{figure}

Such a scenario was recently studied by \cite{LPR12}. It leads to a
few unexpected results.  First, extended emission from higher angles
is very pronounced. This can be understood as a phase space effect:
the average scattering angle is $\approx \Gamma^{-1}$, and $\Gamma$
varies with angle. Thus, more photons that originates from high angles
(with lower $\Gamma$) are observed, compared to the spherical
case. The obtained spectrum for narrow jet ($\theta_j \Gamma_0
\lsim$~few) below the thermal peak is flat ($dN/dE \propto E^{-1}$),
independent of viewing angle, and only weakly dependent on the Lorentz
factor gradient ($p$). A similar result is obtained for wider jets,
observed at $\theta_v \approx \theta_j$, which is the most likely
scenario. The spectral slope calculated in this model is similar to
the average low energy spectral slope observed. For wider jets
($\theta_j \Gamma_0 \gsim$~few), a multicolor black body is
obtained. Second, the high energy spectral slope is modified by a
similar mechanism: as the average scattering angle is $\approx
\Gamma^{-1}$, photons are more likely to diffuse from region of low
$\Gamma$ to region of high $\Gamma$, where they are further
boosted. This leads to a power law spectral slope at high energies,
who's exact shape depends on the assumed jet profile. An example of
the obtained spectra appears in Figure \ref{fig:2}.

\section{Summary}
\label{sec:summary}

In spite of two decades of extensive research, the origin of GRB
prompt emission remains elusive. A renewed interest in understanding
this phenomena occurred with the superb data quality enabled by the {\it
  Fermi} satellite. Following the failure of the synchrotron model, significant
efforts were given to understanding mechanisms that can act to broaden
the Planck spectrum to fit into the observed ``Band'' spectrum.

Three ideas were suggested in recent years: (1) A combined optically
thick and optically thin emission seen simultaneously; (2)
sub-photospheric energy dissipation; and (3) geometrical broadening.
While each of these ideas have its own success, as of today, non of
these provide a full explanation to the observed spectrum.  The
success and weaknesses of any of these ideas are summarized in Table
\ref{tab:1} below. In the table, $\large{(V)}$ represents success,
$\large{(X)}$ represents failure, and $\large{(-)}$ implies that
currently the theory does not contradict the observation, but does not
provide predictions either, or that additional assumptions are
required.

\begin{table}[htbp]
\begin{tabular}{|c|cccc|} 
\hline\hline 
 Key observation & Optically thin & Pure Planck &  Sub phot. energy & Geometrical \\
~ & synchrotron &  + synch. & dissipation & broadening \\
\hline $E_{peak}^{ob} \approx 300$~keV & {\large V} &  {\large V} &  {\large V} &  {\large -V}
\\ 
Narrow $E_{peak}^{ob}$ distribution  & {\large -} &  {\large -} &  {\large V} &  {\large -} \\
$E_{peak} - E_{iso}$ correlation & {\large -} &  {\large X-} &  {\large X-} &  {\large X-} \\
Low energy spectral index & {\Large X} & {\large X} &  {\large -} & {\large V} \\
 $<\alpha> \approx -1$ & ~ & ~ & ~ & ~ \\
\hline
\end{tabular}
\caption{Confronting current theoretical models with key observations.}
\label{tab:1}
\end{table}

Thus, as of today, no single model can fully explain all key
observations, implying plenty of room for new ideas.

\acknowledgements
I would like to thank my collaborator and friend {\bf Felix Ryde} for countless number of useful discussions.

%%-----------------------------
%%      your bibliography
%%-----------------------------
%\bibliographystyle{/Users/apeer/Documents/Bib/apj}
%\bibliographystyle{/Users/apeer/Documents/Bib/astron}
% This points to the .bib database files.
%\bibliography{/Users/apeer/Documents/Bib/abbrevs,/Users/apeer/Documents/Bib/short_abbrevs,/Users/apeer/Documents/Bib/bib_apeer}

%\begin{thebibliography}{99}
%\bibitem[1994]{alref1} Aalto, S. \etal\  1994, A\&A, 286, 365.
%%% Using \cite{Bei} in the text
%\bibitem[1986]{Bei} Beichman, C.A., Neugebauer, G., Habing,
%   H., Clegg, P.E. \& Chester, T.C. 1988, editors, {\it ``IRAS Catalogs and
%   Atlases: Explanatory Supplement''}, NASA RP-1190 (Washington: NASA)
%\bibitem[1987]{ref1987} Beichman, C.A. 1987, ARA\&A, 25, 521
%\bibitem[1987]{so1987} Soifer, B.T., Houck, J.R. and Neugebauer, G. 1987, ARAA,% 25, 187
%\end{thebibliography}

\end{document}